\documentclass{eptcs}
\usepackage{amsmath}
\usepackage{graphicx}
\usepackage{listings}
\usepackage{cite}
\usepackage{array}
\usepackage{courier}

\newcommand{\bm}[1]{\mathbf{#1}}
\newcommand{\Type}[1]{\texttt{#1}}
\newcommand{\defeq}{\stackrel{\mbox{\tiny def}}{=}}

\title{A Simple and Practical Linear Algebra Library Interface with Static Size Checking%
  \thanks{This work was partially supported by JSPS KAKENHI Grant Numbers 22300005, 25540001, 15H02681, and by Mitsubishi Foundation Research Grants in the Natural Sciences.}}
\author{Akinori Abe \qquad\qquad Eijiro Sumii
\institute{Graduate School of Information Sciences\\Tohoku University, Japan}
\email{abe@sf.ecei.tohoku.ac.jp \qquad sumii@sf.ecei.tohoku.ac.jp}}
\def\titlerunning{A Simple and Practical Linear Algebra Library Interface with Static Size Checking}
\def\authorrunning{A. Abe \& E. Sumii}

\begin{document}
\maketitle

\begin{abstract}
Linear algebra is a major field of numerical computation and is widely applied.
Most linear algebra libraries (in most programming languages) do not statically
guarantee consistency of the dimensions of vectors and matrices, causing runtime
errors. While advanced type systems---specifically, dependent
types on natural numbers---can ensure consistency among
the sizes of collections such as lists and arrays
\cite{DBLP:journals/jfp/Xi07,DBLP:conf/frocos/CuiDX05,DBLP:journals/lisp/ChinK01}, such
type systems generally require non-trivial changes to existing
languages and application programs, or tricky type-level programming.

We have developed a linear algebra library interface that verifies
the consistency (with respect to dimensions) of matrix operations
by means of \emph{generative phantom types}, implemented via fairly standard ML types and
module system. To evaluate its usability, we ported to it a practical
machine learning library from a traditional linear algebra library.
We found that most of the changes required for the porting could be
made mechanically, and changes that needed human thought are minor.
\end{abstract}

\section{Introduction}

Linear algebra is a major field of numerical computation and is widely
applied to many domains such as image processing, signal processing, and
machine learning. Some programming languages support built-in matrix operations,
while others are equipped with libraries.
Examples of the former are MatLab,
statistical programming language S, and OptiML
\cite{DBLP:conf/icml/SujeethLBRCWAOO11,OptiML}
(a domain-specific language for machine learning embedded in Scala).
The latter include BLAS \cite{BLAS} and LAPACK \cite{LAPACK},
originally developed for Fortran and now ported to many languages.

Most of the programming languages and linear algebra libraries
do not statically verify consistency of the dimensions of vectors
and matrices. Dimensional inconsistency like
addition of a 3-by-5 matrix and a 5-by-3 matrix
causes runtime errors such as exceptions or, worse, memory corruption.

Advanced type systems---specifically, dependent
types on natural numbers---can statically ensure consistency among
the sizes of collections such as lists and arrays
\cite{DBLP:journals/jfp/Xi07,DBLP:conf/frocos/CuiDX05,DBLP:journals/lisp/ChinK01}.
However, such type systems generally require non-trivial changes to existing
languages and application programs, or tricky type-level programming.

We have developed a linear algebra library interface that guarantees
the consistency (with respect to dimensions) of matrix (and vector)
operations by using \emph{generative phantom types} as fresh
identifiers for statically checking the equality of sizes
(i.e., dimensions). This interface has three attractive features in
particular.

\begin{itemize}
\item It can be implemented only using fairly standard ML types and
  module system.  Indeed, we implemented the interface in OCaml
  (without significant extensions like GADTs) as a wrapper for an
  existing library.
\item For most high-level operations on matrices (e.g., addition
  and multiplication), the consistency of sizes is verified
  statically.  (Certain low-level operations, like accesses to elements
  by indices, need dynamic checks.)
\item Application programs in a
  traditional linear algebra library can be easily migrated to our interface.
  Most of the required changes can be made mechanically.
\end{itemize}

We implemented our static size checking scheme as a
linear algebra library interface---that we call SLAP (Sized Linear Algebra Package,
\url{https://github.com/akabe/slap/})---on top of an existing linear algebra
library LACAML \cite{Lacaml}.
To evaluate the usability of our interface, we ported to it a practical
machine learning library OCaml-GPR \cite{OCaml-GPR} from LACAML,
thereby ensuring the consistency of sizes.

This paper is structured as follows.
In the next section, we explain our basic idea of static size checking
through examples. In Section \ref{sec:typing-BLAS-LAPACK}, we describe
the implementation of our library interface.
In Section \ref{sec:porting}, we report the changes required for the porting of OCaml-GPR
along with the number of lines changed. We compare our approach
with related work in Section \ref{sec:related-works} and conclude
in Section \ref{sec:conclusions}.

\section{Our idea}

Let \lstinline|'n vec| be the type of \lstinline|'n|-dimensional
vectors, \lstinline|('m,'n) mat| be the type of
\lstinline|'m|-by-\lstinline|'n| matrices, and \lstinline|'n size| be
the \emph{singleton type} on natural numbers as sizes of vectors
and matrices, i.e., evaluation of a term of type \lstinline|'n size|
always results in the natural number corresponding to \lstinline|'n|.
The formal type parameters \lstinline|'m| and \lstinline|'n| are instantiated with actual
types that represent the sizes of the vectors or matrices.  Here we only
explain how the dimensions of vectors are represented since those
of matrices (as well as sizes) can be represented similarly.

The abstract type \lstinline|'n vec| can be implemented as any data
type that can represent vectors, e.g., \lstinline|float array|,
where the type parameter \lstinline|'n| is phantom, meaning that it does not
appear on the right hand side of the type definition.  A phantom type
parameter is often instantiated with a type that has no value (i.e., no
constructor), which we call a \emph{phantom type}%
\footnote{In \cite{DBLP:journals/tcs/Blume01}, the term ``phantom type'' has the
same meaning as in this paper. However, it is used differently in some other papers:
types \emph{with} phantom type parameters \cite{DBLP:conf/dsl/LeijenM99}%
---or GADT (generalized algebraic data type) \cite{Hin03Fun}---are called ``phantom types''.}.
The type \lstinline|'n vec| must be made abstract by hiding its implementation
in the module signature so that the size information in the (phantom)
type parameter \lstinline|'n| is not ignored by the typechecker.

It is relatively straightforward to represent dimensions (size
information) as types by, for example, using type-level natural numbers when
this information is decided at compile time.  The main problem is how to represent
dimensions that are unknown until runtime.  It is practically important because
such dynamically determined dimensions are common (e.g., the length of a vector loaded from a file).
Consider the following code for example:
\begin{lstlisting}
let (x : ?${}_1$ vec) = loadvec "file1" in
let (y : ?${}_2$ vec) = loadvec "file2" in
add x y (* add : 'n vec $\to$ 'n vec $\to$ 'n vec *)
\end{lstlisting}
The function \lstinline|loadvec| of type \lstinline|string $\to$ ? vec|
returns a vector of some dimension, loaded from the given path.  The
third line should be ill-typed because the dimensions of \lstinline|x|
and \lstinline|y| are probably different.  (Even if
\lstinline|"file1"| and \lstinline|"file2"| were the same path, the
addition should be ill-typed because the file might have changed between the
two loads.)  Thus, the return type of \lstinline|loadvec| should be
different every time it is called (regardless of the specific values of
the argument).  We call such a return type \emph{generative} because
the function returns a value of a fresh type for each call.

The vector type with generative size information essentially
corresponds to an existentially quantified sized type like
\lstinline|$\exists$n. n vec|.  Our basic idea is to verify
\emph{only} the equality of dimensions by representing them as (only)
generative phantom types.  We implemented this idea in OCaml
and carried out a realistic case study to demonstrate
that it is mostly straightforward to write (or port) application
programs by using our interface (see Section~\ref{sec:porting} for details).

\subsection{Implementation of generative phantom types}

We implement the idea of generative phantom types via packages of types like
\lstinline|$\exists$n. n vec| by using first-class modules and (generative) functors.
For convenience, assume that the abstract type \lstinline|'n vec| is implemented as
\lstinline|float array|.

With first-class modules, the function \lstinline|loadvec| can be defined as
\begin{lstlisting}
module type VEC = sig
  type n            (* corresponds to a generative phantom type `?'. *)
  val value : n vec (* corresponds to `? vec'. *)
end

let loadvec filename = (module struct
                          type n
                          let value = loadarray filename
                        end : VEC)
\end{lstlisting}
where \lstinline|loadarray : string $\to$ float array| loads an array from the given path.
The existential type \lstinline|n| in the module returned by \lstinline|loadvec|
is different every time it is called:
\begin{lstlisting}
module X = (val loadvec "file1" : VEC)
module Y = (val loadvec "file2" : VEC)
add X.value Y.value (* ill-typed since X.n $\ne$ Y.n *)
\end{lstlisting}

Alternatively, this behavior can also be implemented by using
generative functors\footnote{Generative functors have been supported since OCaml 4.02.
Before 4.02, they can be simulated by always passing a module expression of the form
\lstinline|struct ... end| (rather than a named module like \lstinline|M|) to applicative functors.} as follows.
\begin{lstlisting}
module Loadvec (X : sig val filename : string end) () : VEC = struct
  type n
  let value = loadarray filename
end

module X = Loadvec(struct let filename = "file1" end)()
module Y = Loadvec(struct let filename = "file2" end)()
add X.value Y.value (* ill-typed since X.n $\ne$ Y.n *)
\end{lstlisting}
Let us compare the two approaches at the caller site
(i.e., from the viewpoint of a user of our interface) in the code above.
With first-class modules, the user needed to write the signature \lstinline|VEC| for unpacking,
which was not required for generative functors.

In some cases, however, both approaches require more annotations. Specifically,
consider \lstinline|columns : 'm vec list $\to$ ('m,?) mat|,
which creates a matrix by concatenating column vectors in a given list.
The number of rows in the matrix returned by \lstinline|columns| is
equal to the dimension of the given vectors, while the number of columns is
the same as the length of the list, so the latter is generative while the former is not.
Figure \ref{fig:columns} shows implementations and callers of \lstinline|columns| using
first-class modules and a functor.
\begin{figure}
  \begin{minipage}[t]{0.47\textwidth}
    \centering
\begin{lstlisting}[frame=tblr,xrightmargin=10pt,basicstyle={\ttfamily\small}]
(* definition: *)
let columns (type k)
    (l : k vec list) =
  (module struct
     type n and m = k
     let value = ...
   end : MAT with type m = k)



(* caller site: *)
let f (type k) ... =
  let l0 = ... in
  let module X =
    (val concat l0
      : MAT with type m = k)
  in
  ...
\end{lstlisting}
    (a) using first-class modules
  \end{minipage}
  \begin{minipage}[t]{0.52\textwidth}
    \centering
\begin{lstlisting}[frame=tblr,basicstyle={\ttfamily\small}]
(* definition: *)
module Columns (L : sig
      type k
      val l : k vec list
    end) () : MAT with type m = L.k =
struct
  type n and m = L.k
  let value = ...
end

(* caller site: *)
let f (type k') ... =
  let l0 = ... in
  let module X =
    Concat(struct type k = k'
           let l = l0 end)()
  in
  ...
\end{lstlisting}
    (b) using a functor
  \end{minipage}
  \caption{Implementations and callers of \lstinline|columns|}
  \label{fig:columns}
\end{figure}
The signature \lstinline|MAT| in Figure \ref{fig:columns} is defined as:
\begin{lstlisting}
module type MAT = sig
  type m and n
  val value : (m, n) mat
end
\end{lstlisting}
On one hand, when using first-class modules, the signature \lstinline|MAT| of the returned module
\lstinline|columns l0| needs the sharing constraint \lstinline|with type m = k|.
On the other hand, with functors, the actual type argument \lstinline|type k = k'| cannot be omitted.

In summary, the two approaches require different styles of annotations.
We therefore adopted both and often provided two interfaces for the same function so that a user can choose.

\subsection{Free type constructors to represent operations on natural numbers}

Results of some deterministic functions could also be given
generative phantom types but can be given a more precise type using free (in the
sense of free algebra) constructors.
For example, consider the \lstinline|append| function, which
concatenates an $m$-dimensional vector and an $n$-dimensional vector, and returns
a vector of dimension $m+n$.  Theoretically, its typing would be
\begin{lstlisting}
val append : 'm vec $\to$ 'n vec $\to$ ('m + 'n) vec
\end{lstlisting}
if there were addition \lstinline|+| of type-level natural numbers.
Using generative phantom types, however, it can instead be typed as follows
\begin{lstlisting}
val append : 'm vec $\to$ 'n vec $\to$ ? vec
\end{lstlisting}
or, using a type constructor \lstinline|add|,
\begin{lstlisting}
val append : 'm vec $\to$ 'n vec $\to$ ('m, 'n) add vec
\end{lstlisting}
The last alternative can be implemented directly in OCaml (or any language with ML-like parametrized data types)
with a (preferably phantom) type \lstinline|('m,'n) add|, which represents
a size that differs from any other (e.g., \lstinline|('m,'n) add| differs from \lstinline|('n,'m) add|
since the constructor \lstinline|add| is ``free'').
The return type of \lstinline|append|
does not need to be generative because the dimension of the returned vector is
uniquely determined by the dimensions (i.e., the types) of the argument vectors.
The same technique can also be applied for
other functions such as \lstinline|cons| (which adds an element to the head of
a given vector) and \lstinline|tl| (which takes a subvector without the head element.)

\section{Typing of BLAS and LAPACK functions}
\label{sec:typing-BLAS-LAPACK}

BLAS (Basic Linear Algebra Subprograms) \cite{BLAS} and LAPACK (Linear Algebra
PACKage) \cite{LAPACK} are major linear algebra libraries for Fortran.  To evaluate
the effectiveness of our idea, we implemented a linear algebra library
interface as a ``more statically typed'' wrapper of LACAML, which is a BLAS and
LAPACK binding in OCaml.  Our interface is largely similar to
LACAML so that existing application programs can be easily ported.
Here we explain several techniques required for typing the BLAS and
LAPACK functions.

\subsection{Function types that depend on flags}
\label{sec:funtype}

\subsubsection{Transpose flags for matrices}

The BLAS function \lstinline|gemm| multiplies two general matrices:
\begin{lstlisting}
val gemm : ?beta:num_type $\to$ ?c:mat (* $\bm{C}$ *) $\to$
  ?transa:[ `N | `T | `C ] $\to$ ?alpha:num_type $\to$ mat (* $\bm{A}$ *) $\to$
  ?transb:[ `N | `T | `C ] $\to$ mat (* $\bm{B}$ *) $\to$ mat (* $\bm{C}$ *)
\end{lstlisting}
Basically, it executes $\bm{C} := \alpha \bm{A} \bm{B} + \beta \bm{C}$.  The parameters
\lstinline|transa| and \lstinline|transb| specify no transpose
(\lstinline|`N|), transpose (\lstinline|`T|), or conjugate transpose
(\lstinline|`C|) of the matrices $\bm{A}$ and $\bm{B}$ respectively.  For example, if
\lstinline|transa=`N| and \lstinline|transb=`T|, then \lstinline|gemm|
executes $\bm{C} := \alpha \bm{A} \bm{B}^\top + \beta \bm{C}$.  Thus, the \emph{types}
(dimensions) of the matrices change depending on the \emph{values} of
the flags (the transpose of an $m$-by-$n$ matrix is an $n$-by-$m$
matrix).  To implement this behavior, we give each transpose flag a
function type that represents the change in types induced by that
particular transposition, like:
\begin{lstlisting}
type 'a trans (* = [ `N | `T | `C ] *)
val normal : (('m, 'n) mat $\to$ ('m, 'n) mat) trans (* = `N *)
val trans  : (('m, 'n) mat $\to$ ('n, 'm) mat) trans (* = `T *)
val conjtr : (('m, 'n) mat $\to$ ('n, 'm) mat) trans (* = `C *)

val gemm : ?beta:num_type $\to$ ?c:('m, 'n) mat (* $\bm{C}$ *) $\to$
  transa:(('am, 'ak) mat $\to$ ('m, 'k) mat) trans $\to$
  ?alpha:num_type $\to$ ('am, 'ak) mat (* $\bm{A}$ *) $\to$
  transb:(('bk, 'bn) mat $\to$ ('k, 'n) mat) trans $\to$
  ('bk, 'bn) mat (* $\bm{B}$ *) $\to$ ('m, 'n) mat (* $\bm{C}$ *)
\end{lstlisting}
The arguments \lstinline|transa| and \lstinline|transb| are optional in LACAML, but mandatory in our interface
because OCaml restricts the types of parameters to those of the default arguments.
This a shortcoming of the typing of optional arguments in OCaml.

\subsubsection{Side flags for square matrix multiplication}

The BLAS function \lstinline|symm| multiplies a symmetric matrix $\bm{A}$
by a general matrix $\bm{B}$:
\begin{lstlisting}
val symm : ?side:[ `L | `R ] $\to$ ?beta:num_type $\to$
  ?c:mat (* $\bm{C}$ *) $\to$ ?alpha:num_type $\to$ mat (* $\bm{A}$ *) $\to$
  mat (* $\bm{B}$ *) $\to$ mat (* $\bm{C}$ *)
\end{lstlisting}
The parameter \lstinline|side| specifies the ``direction'' of the
multiplication: \lstinline|symm| executes $\bm{C} := \alpha \bm{A} \bm{B} + \beta \bm{C}$
if \lstinline|side| is \lstinline|`L|, and $\bm{C} := \alpha \bm{B} \bm{A} + \beta \bm{C}$
if it is \lstinline|`R|.  If $\bm{B}$ and $\bm{C}$ are $m$-by-$n$ matrices,
$\bm{A}$ is an $m$-by-$m$ matrix in the former case and
$n$-by-$n$ in the latter case.  We implemented these flags as follows:
\begin{lstlisting}
type ('k, 'm, 'n) side (* = [ `L | `R ] *)
val left  : ('m, 'm, 'n) side (* = `L *)
val right : ('n, 'm, 'n) side (* = `R *)
\end{lstlisting}
The parameter \lstinline|'k| in type \lstinline|('k,'m,'n) side|
corresponds to the dimension of the \lstinline|'k|-by-\lstinline|'k|
symmetric matrix $\bm{A}$, and the other parameters \lstinline|'m| and
\lstinline|'n| correspond to the dimensions of the
\lstinline|'m|-by-\lstinline|'n| general matrix $\bm{B}$.  When $\bm{A}$ is
multiplied from the left to $\bm{B}$ (i.e., like $\bm{A}\bm{B}$), \lstinline|'k| is
equal to \lstinline|'m|; therefore, the type of the flag
\lstinline|left| is \lstinline|('m,'m,'n) side|.  Conversely, if $\bm{A}$
is right-multiplied to $\bm{B}$ (i.e., like $\bm{B}\bm{A}$), \lstinline|'k| is equal to
\lstinline|'n|. Thus, the flag \lstinline|right| is given the type
\lstinline|('n,'m,'n) side|.  By using this trick, we can type
\lstinline|symm| as:
\begin{lstlisting}
val symm : side:('k, 'm, 'n) side $\to$ ?beta:num_type $\to$
  ?c:('m, 'n) mat (* $\bm{C}$ *) $\to$ ?alpha:num_type $\to$
  ('k, 'k) mat (* $\bm{A}$ *) $\to$ ('m, 'n) mat (* $\bm{B}$ *) $\to$ ('m, 'n) mat
\end{lstlisting}
The same trick can be applied to other square matrix multiplications as well.

\subsubsection{Flags that change the size of the results}

LAPACK provides routines \lstinline|gesdd| and \lstinline|gesvd|
for singular value decomposition (SVD), a variant of the eigenvalue problem.
It factorizes a given $m$-by-$n$ matrix $\bm{A}$ as
\[
  \bm{A} = \bm{U} \bm{D} \bm{V}^\dagger
\]
where $\bm{U}$ is an $m$-by-$m$ unitary matrix, $\bm{V}^\dagger$ is the conjugate
transpose of an $n$-by-$n$ unitary matrix $\bm{V}$, and $\bm{D}$ is an $m$-by-$n$
matrix with $\min(m,n)$ diagonal elements. The diagonal elements are called singular
values (similar to eigenvalues), and columns of $\bm{U}$ and $\bm{V}$ are called left and
right singular vectors (similar to eigenvectors), respectively.
The singular vectors are sorted according to the corresponding singular values.
The challenge is that \lstinline|gesdd| and \lstinline|gesvd| store
the singular vectors differently to $\bm{U}$, $\bm{V}$, or $\bm{A}$
depending on the flags.

We consider \lstinline|gesdd| first:
\begin{lstlisting}
val gesdd : ?jobz:[ `A | `N | `O | `S ] $\to$
  ?s:vec (* $\bm{D}$ *) $\to$ ?u:mat (* $\bm{U}$ *) $\to$ ?vt:mat (* $\bm{V}^\dagger$ *) $\to$ mat (* $\bm{A}$ *) $\to$
  vec (* $\bm{D}$ *) * mat (* $\bm{U}$ *) * mat (* $\bm{V}^\dagger$ *)
\end{lstlisting}
It computes all singular values but computation of the singular vectors is optional depending on the flags:
\begin{itemize}
\item When the flag \lstinline|jobz| is \lstinline|`A|, all the left and right
  singular values are computed and are stored in \lstinline|u| and \lstinline|vt|.
  In this case, the storage \lstinline|u| and \lstinline|vt| must be an $m$-by-$m$ and an $n$-by-$n$
  matrices, respectively.
\item If \lstinline|jobz| is \lstinline|`S|, only the top $\min(m,n)$ columns in $\bm{U}$
  and the top $\min(m,n)$ rows in $\bm{V}^\dagger$
  are stored in \lstinline|u| and \lstinline|vt|, which
  must be $m$-by-$\min(m,n)$ and $\min(m,n)$-by-$n$, respectively.
\item Flag \lstinline|`O| specifies to overwrite the matrix $\bm{A}$ with
  the top singular vectors as follows:
  \begin{itemize}
  \item If $m \ge n$, $\bm{A}$ is overwritten with the top $\min(m,n)$ columns of
    $\bm{U}$, and the $n$ rows of $\bm{V}^\dagger$ are returned in \lstinline|vt|;
    thus \lstinline|vt| is $n$-by-$n$ while \lstinline|u| is not used.
  \item If $m < n$, $\bm{A}$ is overwritten with the top $\min(m,n)$ rows of
    $\bm{V}^\dagger$, and the $m$ columns of $\bm{U}$ are returned in \lstinline|u|,
    so \lstinline|u| is $m$-by-$m$ and \lstinline|vt| is not used.
  \end{itemize}
\item For \lstinline|`N|, no singular vectors are calculated at all
(only singular values are returned).
\end{itemize}
We implement the dependency of the sizes of \lstinline|u| and
\lstinline|vt| only on the value of \lstinline|jobz|, ignoring
whether $m \ge n$ or $m < n$ in the case of \lstinline|`O|
(i.e., \lstinline|u| is required to be $m$-by-$m$ and \lstinline|vt| to be $n$-by-$n$
whenever the flag is \lstinline|`O| even though only one of them is used;
in addition, the SLAP-version of \lstinline|gesdd| returns
a value of type \lstinline|_ vec * _ mat option * _ mat option| instead of
\lstinline|_ vec * _ mat * _ mat| to avoid allocating dummy matrices
when $\bm{U}$ or $\bm{V}$ are not returned).

Based on the ideas above,
we defined the SVD job flags and the type of \lstinline|gesdd| as follows:
\begin{lstlisting}
type ('a, 'b, 'c, 'd, 'e) svd_job (* = [ `A | `N | `O | `S ] *)

val svd_all       : ('a, 'a,  _,  _,  _) svd_job (* = `A *)
val svd_top       : ('a,  _, 'a,  _,  _) svd_job (* = `S *)
val svd_overwrite : ('a,  _,  _, 'a,  _) svd_job (* = `O *)
val svd_no        : ('a,  _,  _,  _, 'a) svd_job (* = `N *)

val gesdd : jobz:('u_cols * 'vt_rows,
                  'm * 'n,
                  ('m, 'n) min * ('m, 'n) min,
                  'm * 'n,
                  z * z)
                 svd_job $\to$
            ?s:(('m, 'n) min, cnt) vec        (* $\bm{D}$ *) $\to$
            ?u:('m, 'u_cols, 'u_cd) mat       (* $\bm{U}$ *) $\to$
            ?vt:('vt_rows, 'n, 'vt_cd) mat    (* $\bm{V}^\dagger$ *) $\to$
            ('m, 'n, 'a_cd) mat               (* $\bm{A}$ *) $\to$
              (('m, 'n) min, 's_cd) vec         (* $\bm{D}$ *) *
              ('m, 'u_cols, 'u_cd) mat option   (* $\bm{U}$ *) *
              ('vt_rows, 'n, 'vt_cd) mat option (* $\bm{V}^\dagger$ *)
\end{lstlisting}
In the type of \lstinline|svd_all|, the first type parameter is the
same as the second. When it is passed to \lstinline|jobz|,
the OCaml typechecker unifies \lstinline|'u_cols * 'vt_rows| and
\lstinline|'m * 'n| so that \lstinline|u| and \lstinline|vt| have types
\lstinline|('m,'m) mat| and \lstinline|('n,'n) mat|, respectively.
Similarly, if \lstinline|svd_top| is specified, \lstinline|u| and
\lstinline|vt| are typed as \lstinline|('m,('m,'n) min) mat| and
\lstinline|(('m,'n) min,'n) mat|. In the case of \lstinline|svd_overwrite|,
\lstinline|u : ('m,'m) mat| and \lstinline|vt : ('n,'n) mat| are derived.
For \lstinline|svd_no|, \lstinline|u| and \lstinline|vt| have
types \lstinline|('m,z) mat| and \lstinline|(z,'n) mat|,
where \lstinline|z| is a nullary type constructor representing $0$.
(These types
are virtually equal to \lstinline|(z,z) mat| because a matrix of
\lstinline|('m,z) mat| or \lstinline|(z,'n) mat| has no element at all.)

Second, we consider the type of \lstinline|gesvd|.
The original typing (in LACAML) of \lstinline|gesvd| is:
\begin{lstlisting}
val gesvd : ?jobu:[ `A | `N | `O | `S ] $\to$ ?jobvt:[ `A | `N | `O | `S ] $\to$
  ?s:vec (* $\bm{D}$ *) $\to$ ?u:mat (* $\bm{U}$ *) $\to$ ?vt:mat (* $\bm{V}^\dagger$ *) $\to$ mat (* $\bm{A}$ *) $\to$
  vec (* $\bm{D}$ *) * mat (* $\bm{U}$ *) * mat (* $\bm{V}^\dagger$ *)
\end{lstlisting}
It takes two SVD job flags \lstinline|jobu| and \lstinline|jobvt|
for the computation of the singular vectors in $\bm{U}$ and $\bm{V}^\dagger$.
When \lstinline|jobu| is \lstinline|`A|, \lstinline|`S|, or \lstinline|`N|,
all, the top $\min(m,n)$, or no columns in $\bm{U}$ are computed respectively.
If it is \lstinline|`O|, \lstinline|a| is overwritten with the top $\min(m,n)$ columns.
The meaning of \lstinline|jobvt| is similar.
(It is a runtime error to give \lstinline|`O| for both \lstinline|jobu| and \lstinline|jobvt|,
because $\bm{A}$ cannot accommodate $\bm{U}$ and $\bm{V}^\dagger$ at the same time.)
The type of \lstinline|gesvd| can be given using the above definition
of SVD flags:
\begin{lstlisting}
val gesvd : jobu:('u_cols, 'm, ('m, 'n) min, ('m, 'n) min, z) svd_job $\to$
            jobvt:('vt_cols, 'm, ('m, 'n) min, ('m, 'n) min, z) svd_job $\to$
            ?s:(('m, 'n) min, cnt) vec     (* $\bm{D}$ *) $\to$
            ?u:('m, 'u_cols, 'u_cd) mat    (* $\bm{U}$ *) $\to$
            ?vt:('vt_rows, 'n, 'vt_cd) mat (* $\bm{V}^\dagger$ *) $\to$
            ('m, 'n, 'a_cd) mat            (* $\bm{A}$ *) $\to$
              (('m, 'n) min, 's_cd) vec      (* $\bm{D}$ *) *
              ('m, 'u_cols, 'u_cd) mat       (* $\bm{U}$ *) *
              ('vt_rows, 'n, 'vt_cd) mat     (* $\bm{V}^\dagger$ *)
\end{lstlisting}
In accordance with the type-level trick described in the next section,
$\bm{D}$ of \lstinline|gesdd| and \lstinline|gesvd|
has type \lstinline|(('m, 'n) min, cnt) vec| in the argument (contravariant position)
and type \lstinline|(('m, 'n) min, 's_cd) vec| in the return value (covariant position)
because $\bm{D}$ refers to a contiguous memory region.

\subsection{Subtyping for discrete memory access}
\label{discrete}

In Fortran, elements of a matrix are stored in column-major order in a
flat, contiguous memory region.  BLAS and LAPACK functions can take
part of a matrix (like a row, a column, or a submatrix) and use it for
computation without copying the elements, so they need to access the
memory discretely in order to access the elements.  However, some
original functions of LACAML do not support such discrete access.
For compatibility and soundness, we need to prevent those
functions from receiving (sub)matrices that require discrete accesses
while allowing the converse (i.e., the other functions may receive contiguous
matrices as well as discrete ones).  We achieved this by
extending the type definition of matrices by adding a third
parameter for ``contiguous or discrete'' flags (in addition to the
existing two parameters for dimensions):
\begin{lstlisting}
type ('m, 'n, 'cnt_or_dsc) mat (* 'm-by-'n matrices *)
type cnt (* phantom *)
type dsc (* phantom *)
\end{lstlisting}
Then, formal arguments that may be either contiguous or discrete
matrices are given type \lstinline|('m,'n,'cnt_or_dsc) mat|, while the
types of (formal) arguments that \emph{must} be contiguous are
specified in the form \lstinline|('m,'n,cnt) mat|.  In contrast,
return values that may be either contiguous or discrete have type
\lstinline|('m,'n,dsc) mat|, while those that are known to be always
contiguous are typed \lstinline|('m,'n,'cnt_or_dsc) mat| so that they
can be mixed with discrete matrices.

Appendix \ref{subtyping} presents a generalization of the idea in this
subsection to encode subtyping via phantom types.

\subsection{Dynamic checks remaining}

Although many high-level operations provided by BLAS and LAPACK can be
statically verified by using the scheme described above as far as equalities
of dimensions are concerned, other operations still require dynamic checks
for inequalities:
\begin{itemize}
\item \lstinline|get| and \lstinline|set| operations allow accesses to
  an element of a matrix via given indices, which must be less than
  the dimensions of the matrix.
  This inequality is dynamically checked and
  no static size constraint is imposed.
  (The use of these low-level functions is therefore unrecommended
  and high-level matrix operations such as \lstinline|map| or BLAS/LAPACK functions
  should be used instead; see Section \ref{sec:manual-changes} for details.)
\item As mentioned above, BLAS and LAPACK functions can take a
  submatrix without copying it.  Our original function \lstinline|submat|
  returns such a submatrix for the dimensions given. This submatrix must be smaller than
  the original matrix.
\item An efficient memory arrangement for band matrices in BLAS and
  LAPACK, called the \emph{band storage scheme}, requires a dynamic
  check that the specified numbers of sub- and super-diagonals are
  smaller than the size of the band matrix; for details, see Section
  \ref{band}.
\item There are several high-level functions with specifications that
  essentially involve submatrices and inequalities of indices, such as
  \lstinline|syevr| (for finding eigenvalues),
  \lstinline|orgqr|, and \lstinline|ormqr| (for QR factorization).
\item For most LAPACK functions, the workspace for computation can be
  given as an argument. It must be larger than the minimum
  workspace required as determined by each function (and other
  arguments).
\end{itemize}

We gave these dynamically checked functions the suffix \lstinline|_dyn|, like
\lstinline|get_dyn| and \lstinline|set_dyn| (with the exception of the
last bullet point since almost \emph{all} LAPACK functions can take the
workspace as a parameter whose size must be checked dynamically).

These inequalities are dynamically checked by our library interface
for the sake of usability and compatibility with LACAML because the interface would become
too complex if the inequalities were represented as types: for example, consider
\lstinline|('m,'n) le| as a type that would represent
\lstinline|'m $\le$ 'n|; then the types of
\lstinline|get| and \lstinline|set| could be given as follows
(\lstinline|one| is the type for the natural number $1$):
\begin{lstlisting}
val get : (one, 'i) le $\to$ ('i, 'n) le $\to$ ('n, _) vec $\to$ 'i size $\to$ float
val set : (one, 'i) le $\to$ ('i, 'n) le $\to$ ('n, _) vec $\to$ 'i size $\to$ float $\to$ unit
\end{lstlisting}
Not only the users must pass two extra arguments, but they would also
have to \emph{derive} the inequalities by applying functions for
axioms such as reflexivity and transitivity.
We have rejected this approach because, after all, users want to
write linear algebra, not proof terms.

\subsection{Inequality capabilities}

Although most inequalities are dynamically checked,
we introduced capabilities and types for guaranteeing inequalities
in the following two cases as they arise naturally.

\subsubsection{Submatrices and subvectors}
\label{sec:submatrix}

All BLAS and LAPACK functions support operations on subvectors and submatrices.
For instance,
\begin{lstlisting}
lange ~m ~n ~ar ~ac a
\end{lstlisting}
computes the norm of the $m$-by-$n$ submatrix in matrix \lstinline|a| where
element $(i,j)$ corresponds to the $(i+\mathit{ar}-1, j+\mathit{ac}-1)$
element of \lstinline|a|\footnote{The actual \lstinline|lange| function takes
another parameter \lstinline|norm| that indicates the kind of
the norm, e.g., 1-norm, Frobenius norm, etc.}.
We cannot statically verify whether this function
call is safe, i.e., the submatrix is indeed a submatrix of \lstinline|a|.
Since adding dynamic checks to \emph{all} BLAS and LAPACK functions is undesirable,
we have introduced a new, separate function
\begin{lstlisting}
val submat_dyn : 'm size $\to$ 'n size $\to$ ?ar:int $\to$ ?ac:int $\to$
                 ('k, 'l, 'cnt_or_dsc) mat $\to$ ('m, 'n, dsc) mat
\end{lstlisting}
to return a submatrix of the given matrix. Then the type of \lstinline|lange| can be given simply as:
\begin{lstlisting}
val lange : ('m, 'n, 'cnt_or_dsc) mat $\to$ float
\end{lstlisting}
Thus \lstinline|lange| itself requires no dynamic checks because the inequalities are already
checked by \lstinline|submat_dyn|.

Similarly, we defined the function \lstinline|subvec_dyn| to return a subvector of a given vector or matrix.

\subsubsection{Band storage scheme}
\label{band}

\begin{figure}
  \centering
  \begin{tabular}{>{\centering\arraybackslash}m{0.5\textwidth}>{\centering\arraybackslash}m{0.5\textwidth}}
    \includegraphics{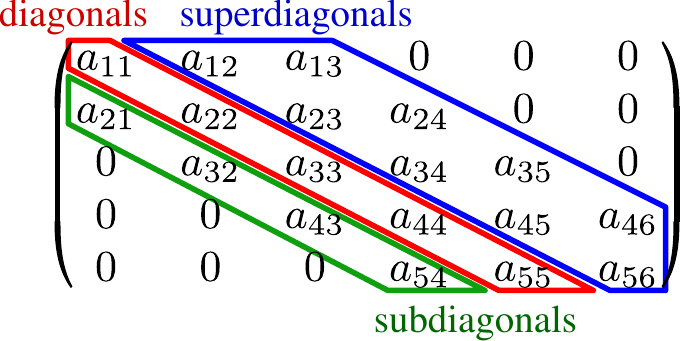} &
    \includegraphics{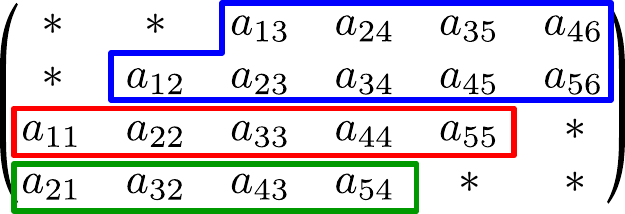}\vfill \\
    (a) Band matrix & (b) Band-storage representation
  \end{tabular}
  \caption{Band storage scheme}
  \label{fig:band-storage}
\end{figure}

In BLAS (and therefore LAPACK), an $m$-by-$n$ band matrix with $\mathit{kl}$ subdiagonals and
$\mathit{ku}$ superdiagonals can be stored in a matrix with $\mathit{kl}+\mathit{ku}+1$ rows and
$n$ columns. This \emph{band storage scheme} is used in practice only when $\mathit{kl}, \mathit{ku} \ll \min(m, n)$.
The $(i,j)$ element in the original matrix is stored in the $(\mathit{ku}+1+i-j, j)$ element of the
band-storage representation, where $\max(1, j-\mathit{ku}) \le i \le \min(m, j+\mathit{kl})$.
Figure \ref{fig:band-storage} shows a 5-by-6 band matrix with two superdiagonals and one subdiagonal,
and its band-storage representation as an example (the $*$ symbol denotes an
unused element).
A matrix like (b) is passed to special functions like \lstinline|gbmv| (which multiplies
a band matrix to a vector) for band-storage representations.

We implemented our original function
\lstinline|geband_dyn|\footnote{\lstinline|geband_dyn| means GEneral BAND matrix.
There is also a variant \lstinline|syband_dyn| for symmetric band matrices, hence the name.}
that converts a band matrix into band-storage representation
\begin{lstlisting}
val geband_dyn : 'kl size $\to$ 'ku size $\to$ ('m, 'n) mat $\to$
                 (('m, 'n, 'kl, 'ku) geband, 'n) mat
\end{lstlisting}
where the phantom type \lstinline|('m,'n,'kl,'ku) geband|
guarantees the inequalities $\mathit{kl} < m$ and $\mathit{ku} < n$,
and represents the height of the band-storage representation
of an \lstinline|'m|-by-\lstinline|'n| band matrix with \lstinline|'kl| subdiagonals
and \lstinline|'ku| superdiagonals.
\lstinline|geband_dyn| dynamically checks
the inequalities and performs the conversion.

The matrix-vector multiplication function \lstinline|gbmv| in LACAML\footnote{We
omit an optional argument \lstinline|?n:int| to take a submatrix of $\bm{A}$,
since we separate such submatrix operations as in Section~\ref{sec:submatrix}.} is typed as
\begin{lstlisting}
val gbmv : ?m:int $\to$ ?beta:num_type $\to$
  ?y:vec $\to$ (* $\bm{y}$ *)
  ?trans:trans3 $\to$
  ?alpha:num_type $\to$
  mat (* $\bm{A}$ *) $\to$
  int (* $\mathit{kl}$ *) $\to$ int (* $\mathit{ku}$ *) $\to$ vec (* $\bm{x}$ *) $\to$ vec (* $\bm{y}$ *)
\end{lstlisting}
while we give it the following type using \lstinline|('m,'n,'kl,'ku) geband|.
\begin{lstlisting}
val gbmv : m:'a_m size $\to$ ?beta:num_type $\to$
  ?y:'m vec (* $\bm{y}$ *) $\to$
  trans:(('a_m, 'a_n) mat $\to$ ('m, 'n) mat) trans $\to$
  ?alpha:num_type $\to$
  (('a_m, 'a_n, 'kl, 'ku) geband, 'a_n) mat (* $\bm{A}$ *) $\to$
  'kl size (* $\mathit{kl}$ *) $\to$ 'ku size (* $\mathit{kl}$ *) $\to$ 'n vec (* $\bm{x}$ *) $\to$ 'm vec (* $\bm{y}$ *)
\end{lstlisting}
This function computes $\bm{y} := \alpha \bm{A} \bm{x} + \beta \bm{y}$ or
$\bm{y} := \alpha \bm{A}^\top \bm{x} + \beta \bm{y}$ for an $m$-by-$n$ matrix $\bm{A}$,
an $n$-dimensional vector $\bm{x}$, and an $m$-dimensional vector $\bm{y}$.
Note that \lstinline|gbmv| requires no dynamic check of $\mathit{kl} < m$ or $\mathit{ku} < n$
because the inequalities are statically guaranteed by the type \lstinline|('m,'n,'kl,'ku) geband|.

\section{Porting of OCaml-GPR}
\label{sec:porting}

To evaluate the usability of SLAP, we ported OCaml-GPR---a practical machine learning
library for Gaussian process regression, written using LACAML
by the same author---to use SLAP.
The ported library, called SGPR, is available at \url{https://github.com/akabe/sgpr}.

Just to give a (very) rough feel for the library via a simple (but
non-trivial) example, the type of a function that ``calculates a
covariance matrix of inputs given a kernel''
\begin{lstlisting}
val calc_upper : Kernel.t $\to$ Inputs.t $\to$ mat
\end{lstlisting}
is augmented like
\begin{lstlisting}
val calc_upper : ('D, _, _) Kernel.t $\to$ ('D, 'n) Inputs.t $\to$
                 ('n, 'n, 'cnt_or_dsc) mat
\end{lstlisting}
indicating that it takes \lstinline|'n| vectors of dimension
\lstinline|'D| and returns an \lstinline|'n|-by-\lstinline|'n|
contiguous matrix.

We added comments of the form \lstinline|(*! $\mathit{label}$ *)| on
lines changed in the SGPR source code to investigate the categories
and numbers of changes required for the porting, where
$\mathit{label}$ corresponds to a category of changes, e.g.,
\lstinline|(*! IDX *)| for replacement of index-based accesses.  We
classified the changes into 19 categories, some of which we outline
here (see \url{https://akabe.github.io/sgpr/changes.pdf} for the
others).

\subsection{Changes that could be made mechanically}

Of the 19 categories of changes required, 12 could be made
mechanically.  Here we describe three representative examples.

\paragraph*{Replacement of index-based accesses (IDX)}
In LACAML, the syntax sugar \lstinline|x.{i,j}| can be used for
index-based accesses to elements of vectors and matrices because they
are implemented with the built-in OCaml module \lstinline|Bigarray|.
In SLAP, however, this syntax sugar cannot be used since
\lstinline|('n,'cd) vec| and \lstinline|('m,'n,'cd) mat| are abstract%
\footnote{Since the development version of OCaml (4.03, not released yet) will support
user-defined index operators \lstinline|.\{\}| and \lstinline|.\{,\}|,
our replacement functions \lstinline|get_dyn| and \lstinline|set_dyn| will be unnecessary in the (near) future.};
one must use the \lstinline|get_dyn| and \lstinline|set_dyn|
functions instead.

\paragraph*{Rewriting of the flags (RF)}
The transpose flags had to be rewritten from \lstinline|`N|,
\lstinline|`T|, and \lstinline|`C| to \lstinline|normal|,
\lstinline|trans|, and \lstinline|conjtr| (and similarly for side and
SVD job flags) for the sake of typing as described in Section
\ref{sec:funtype}.

\paragraph*{Insertion of type parameters (ITP)}
Recall that we changed the types \lstinline|vec| and \lstinline|mat|
on the right hand side of a type definition to \lstinline|('n,'cd) vec|
and \lstinline|('m,'n,'cd) mat|, respectively.  It is then
necessary to add the type parameters \lstinline|'m|, \lstinline|'n|,
and \lstinline|'cd| on the left hand side as well.  Theoretically, it
suffices to give fresh type parameters to all \lstinline|vec| and
\lstinline|mat|. For example\footnote{This example is imaginary and
  \emph{not} from the real OCaml-GPR code, which is too complex to
  explain on paper.},
\begin{lstlisting}
module M : sig
  type t
  val f : int $\to$ t
end = struct
  type t = {
    n : int;
    id : mat;
  }
  let f n = { n; id = Mat.identity n }
end
\end{lstlisting}
should be rewritten:
\begin{lstlisting}
module M : sig
  type ('a, 'b, 'c, 'd) t
  val f : 'a size $\to$ ('a, 'a, 'a, 'cnt_or_dsc) t
end = struct
  type ('a, 'b, 'c, 'd) t = {
    n : 'a size;
    id : ('b, 'c, 'd) mat;
  }
  let f n = { n; id = Mat.identity n }
end
\end{lstlisting}
Note that, in the latter code, type parameters of function \lstinline|f| could be
automatically inferred by OCaml.

In practice, however, it would introduce too many type parameters
to take \emph{all} of them fresh.  We thus unified the type parameters
that are known to be always equal by looking at the constructor functions such as \lstinline|f|:
\begin{lstlisting}
module M : sig
  type ('n, 'cnt_or_dsc) t (*! ITP *)
  val f : 'n size $\to$ ('n, 'cnt_or_dsc) t (*! ITP *)
end = struct
  type ('n, 'cnt_or_dsc) t = { (*! ITP *)
    n : 'n size; (*! ITP *)
    id : ('n, 'n, 'cnt_or_dsc) mat; (*! ITP *)
  }
  let f n = { n; id = Mat.identity n }
end
\end{lstlisting}

\subsection{Changes that had to be made manually}
\label{sec:manual-changes}

Other changes needed human brain and had to be made manually.  We here
explain the following two representatives of the seven categories of manual
changes. (The other categories are function types that depend on the
values of arguments, and some ad hoc changes; again see
\url{https://akabe.github.io/sgpr/changes.pdf}.)

\paragraph{Insertion of type annotations (ITA)}
When a matrix operation is implemented by using low-level index-based
access functions, its size constraints cannot be inferred statically
(since they are checked only at runtime).  For example, consider the
function \lstinline|axby|, which calculates $\alpha \bm{x} + \beta
\bm{y}$ where $\alpha$ and $\beta$ are scalar values, and $\bm{x}$ and
$\bm{y}$ are vectors.
\begin{lstlisting}
let axby alpha x beta y =
  let n = Vec.dim x in
  let z = Vec.create n in
  for i = 1 to Size.to_int n do
    Vec.set_dyn z i
      (alpha *. (Vec.get_dyn x i) +. beta *. (Vec.get_dyn y i))
  done;
  z
\end{lstlisting}
The dimensions of vectors $\bm{x}$ and $\bm{y}$ must be the same,
but OCaml cannot infer that:
\begin{lstlisting}
val axby : float $\to$ ('n, _) vec $\to$ float $\to$ ('m, _) vec $\to$ ('n, _) vec
\end{lstlisting}
There are two ways to solve this problem.
One is to type-annotate the function by hand:
\begin{lstlisting}
let axby alpha (x : ('n, _) vec) beta (y : ('n, _) vec) =
  ...
\end{lstlisting}
The other is to use high-level operations such as BLAS/LAPACK functions or
\lstinline|map2| instead of the low-level operations \lstinline|get_dyn|
and \lstinline|set_dyn|:
\begin{lstlisting}
let axby alpha x beta y =
  let z = copy y in (* z := y *)
  scal beta z;      (* z := beta * z *)
  axpy ~alpha ~x z; (* z := alpha * x + z *)
  z
\end{lstlisting}
or
\begin{lstlisting}
let axby alpha x beta y =
  Vec.map2 (fun xi yi $\to$ alpha *. xi +. beta *. yi) x y
\end{lstlisting}
In either way, we need to rewrite existing programs by considering their meanings.

We encountered five such functions in OCaml-GPR and adopted the former approach for all of them
so as to keep the changes minimal.

\paragraph{Escaping generative phantom types (EGPT)}
We needed to prevent a generative phantom type from escaping its scope.
Consider the following function implemented using LACAML\footnote{Again,
this is just an example for pedagogy;
SLAP itself supplies a similar function \lstinline|Vec.of_array|.}.
\begin{lstlisting}
let vec_of_array a = Vec.init (Array.length a) (fun i $\to$ a.(i))
\end{lstlisting}
It converts an array into a vector. In SLAP, the above code causes a type error
because \lstinline|Vec.init| expects a size value of singleton type
\lstinline|'n size|, not an integer, as the first argument.
We thus should convert the integer into a size with
\lstinline|Size.of_int_dyn : int $\to$ (module SIZE)|\footnote{
It raises an exception if the given integer is negative because a size
must be non-negative, hence the suffix \lstinline|_dyn|.} as follows:
\begin{lstlisting}
module type SIZE = sig
  type n
  val value : n size
end

let vec_of_array a =
  let module N = (val Size.of_int_dyn (Array.length a) : SIZE) in
  Vec.init N.value (fun i $\to$ a.(i))
\end{lstlisting}
The fix may seem correct, but it does not type-check in OCaml because the
generative phantom type \lstinline|N.n| escapes its scope.

There are two ways to handle this situation in SLAP.
One is to insert the argument
\lstinline|n| for the size of the array and remove the generative phantom
type from the function:
\begin{lstlisting}
let vec_of_array n a = (* : 'n size $\to$ float array $\to$ ('n, _) vec *)
  if Size.to_int n <> Array.length a then invalid_arg "error";
  Vec.init n (fun i $\to$ a.(i))
\end{lstlisting}
In this approach, the generative phantom type should be given from the outside of the function.
For example, \lstinline|vec_of_array| can be called like:
\begin{lstlisting}
let f a =
  let module N = (val Size.of_int_dyn (Array.length a) : SIZE) in
  let v = vec_of_array N.value a in
  printf "%a" pp_rfvec v
\end{lstlisting}
However, the dynamic check in the definition of \lstinline|vec_of_array| is redundant
in this case.

The other way is to define \lstinline|vec_of_array : float array $\to$ (?, _) vec|
using a first-class module (or a functor):
\begin{lstlisting}
let vec_of_array a = (* : float array $\to$ (module VEC) *)
  let module N = (val Size.of_int_dyn (Array.length a) : SIZE) in
  (module struct
     type n = N.n
     let value = Vec.init N.value (fun i $\to$ a.(i))
   end : VEC)
\end{lstlisting}
In this approach, the generative phantom type is created inside the function.
No extra dynamic check is required, but the user needs to write a type
annotation for the returned module like \lstinline|(val vec_of_array a : VEC)| at the caller site.
We suppose that conversion from the first \lstinline|vec_of_array|
implemented in LACAML into the last code can be made automatically
by inserting packing and unpacking of first-class modules:
packing should be inserted where a generative phantom type escapes, and
unpacking should be inserted where the contents (a vector or a matrix) of the packed module are used.
It is however burdensome to manually insert them
because of the heavy syntax.
We thus adopted the former approach for our (manual) porting.

Note that, in either approach, the conversion may introduce another escape of
the generative phantom type at the caller site and therefore may have to be repeated
(until it reaches the main routine in the worst case, though we conjecture
from our experiences with SLAP that such cases are rare).

\subsection{Results}

Table \ref{tbl:mechanical} shows the number of lines that required
mechanical changes.
The major change was ITP (6.17 \%) because OCaml-GPR consists
of several large modules with a large number of functions involving sized types.
Most of the ITP changes have been made in lib/interfaces.ml, which defines all
the signatures provided by OCaml-GPR.
IDX was the second largest (3.56 \%) because index-based access functions
are frequently used in OCaml-GPR,
even when they could be replaced with high-level matrix operations
such as \lstinline|map|.
\begin{table}
  \caption{Number and percentage of mechanically changed lines}
  \label{tbl:mechanical}
  \centering
  {\footnotesize\begin{tabular}{lrrrrrrrrrrrrr}
  \hline
  & S2I & SC & SOP & I2S & IDX & RF & IF & SUB & ETA & RID & RMDC & ITP & \textbf{Total} \\
  \hline
  lib/block\_diag.mli & 0 & 0 & 0 & 0 & 0 & 0 & 0 & 0 & 0 & 1 & 0 & 5 & 6 \\
  lib/block\_diag.ml & 1 & 0 & 0 & 0 & 0 & 0 & 0 & 0 & 0 & 1 & 6 & 1 & 9 \\
  lib/cov\_const.mli & 0 & 0 & 0 & 0 & 0 & 0 & 0 & 0 & 0 & 0 & 0 & 5 & 5 \\
  lib/cov\_const.ml & 2 & 0 & 0 & 0 & 1 & 0 & 0 & 0 & 0 & 2 & 0 & 9 & 14 \\
  lib/cov\_lin\_one.mli & 0 & 0 & 0 & 0 & 0 & 0 & 0 & 0 & 0 & 1 & 0 & 5 & 6 \\
  lib/cov\_lin\_one.ml & 0 & 0 & 0 & 0 & 1 & 4 & 2 & 0 & 0 & 2 & 0 & 9 & 17 \\
  lib/cov\_lin\_ard.mli & 0 & 0 & 0 & 0 & 0 & 0 & 0 & 0 & 0 & 1 & 0 & 5 & 6 \\
  lib/cov\_lin\_ard.ml & 7 & 0 & 0 & 0 & 10 & 5 & 2 & 0 & 0 & 2 & 0 & 9 & 32 \\
  lib/cov\_se\_iso.mli & 0 & 0 & 0 & 0 & 0 & 0 & 0 & 0 & 0 & 1 & 0 & 5 & 6 \\
  lib/cov\_se\_iso.ml & 18 & 4 & 0 & 0 & 31 & 0 & 0 & 0 & 0 & 2 & 2 & 14 & 71 \\
  lib/cov\_se\_fat.mli & 0 & 0 & 0 & 0 & 0 & 0 & 0 & 0 & 0 & 1 & 0 & 10 & 11 \\
  lib/cov\_se\_fat.ml & 43 & 9 & 1 & 0 & 87 & 2 & 2 & 0 & 0 & 2 & 8 & 23 & 174 \\
  lib/fitc\_gp.mli & 0 & 0 & 0 & 0 & 0 & 0 & 0 & 0 & 0 & 0 & 0 & 0 & 0 \\
  lib/fitc\_gp.ml & 81 & 3 & 3 & 0 & 63 & 19 & 26 & 14 & 34 & 3 & 15 & 69 & 298 \\
  lib/interfaces.ml & 0 & 0 & 0 & 0 & 0 & 0 & 0 & 0 & 0 & 1 & 0 & 196 & 197 \\
  lib/gpr\_utils.ml & 10 & 0 & 0 & 0 & 13 & 0 & 2 & 0 & 0 & 4 & 17 & 1 & 46 \\
  app/ocaml\_gpr.ml & 13 & 0 & 2 & 4 & 10 & 0 & 0 & 0 & 0 & 3 & 0 & 8 & 35 \\
  \hline
  \textbf{Total} & 175 & 16 & 6 & 4 & 216 & 30 & 34 & 14 & 34 & 27 & 48 & 374 & 933 \\
  \textbf{Percentage} & 2.89 & 0.26 & 0.10 & 0.07 & 3.56 & 0.49 & 0.56 & 0.23 & 0.56 & 0.45 & 0.79 & 6.17 & 15.39 \\
  \hline
  \end{tabular}
}
\end{table}

Table \ref{tbl:manual} shows the numbers and percentages of lines
for which the required changes had to be made manually, and
Table \ref{tbl:all} gives the total of all changes.
Overall, 18.4 \% of lines required some changes, out of which (with some overlap)
15.4 \% were mechanical and 3.6 \% required human brain.
From these results, we conjecture in general that
the number of non-trivial changes required for a user program of SLAP is small.
\begin{table}
  \caption{Number and percentage of manually changed lines}
  \label{tbl:manual}
  \centering
  {\footnotesize\begin{tabular}{lrrrrrrrr}
  \hline
  & ITA & EGPT & O2L & FT & ET & DKS & FS & \textbf{Total} \\
  \hline
  lib/block\_diag.mli & 0 & 0 & 0 & 0 & 0 & 0 & 0 & 0 \\
  lib/block\_diag.ml & 0 & 0 & 0 & 0 & 0 & 0 & 0 & 0 \\
  lib/cov\_const.mli & 0 & 0 & 0 & 0 & 0 & 0 & 2 & 2 \\
  lib/cov\_const.ml & 0 & 1 & 0 & 0 & 0 & 1 & 8 & 10 \\
  lib/cov\_lin\_one.mli & 0 & 0 & 0 & 0 & 0 & 0 & 2 & 2 \\
  lib/cov\_lin\_one.ml & 0 & 0 & 0 & 0 & 0 & 1 & 12 & 13 \\
  lib/cov\_lin\_ard.mli & 0 & 0 & 0 & 0 & 0 & 0 & 2 & 2 \\
  lib/cov\_lin\_ard.ml & 0 & 0 & 0 & 0 & 0 & 1 & 8 & 9 \\
  lib/cov\_se\_iso.mli & 0 & 0 & 0 & 0 & 0 & 0 & 2 & 2 \\
  lib/cov\_se\_iso.ml & 2 & 0 & 0 & 0 & 0 & 1 & 6 & 9 \\
  lib/cov\_se\_fat.mli & 0 & 0 & 0 & 4 & 0 & 0 & 0 & 4 \\
  lib/cov\_se\_fat.ml & 0 & 0 & 0 & 28 & 0 & 1 & 1 & 30 \\
  lib/fitc\_gp.mli & 0 & 0 & 0 & 0 & 0 & 0 & 0 & 0 \\
  lib/fitc\_gp.ml & 0 & 28 & 31 & 16 & 0 & 0 & 0 & 68 \\
  lib/interfaces.ml & 0 & 11 & 7 & 5 & 0 & 3 & 0 & 26 \\
  lib/gpr\_utils.ml & 6 & 0 & 0 & 0 & 0 & 0 & 1 & 7 \\
  app/ocaml\_gpr.ml & 0 & 17 & 0 & 6 & 16 & 0 & 0 & 35 \\
  \hline
  \textbf{Total} & 8 & 57 & 38 & 59 & 16 & 8 & 44 & 219 \\
  \textbf{Percentage} & 0.13 & 0.94 & 0.63 & 0.97 & 0.26 & 0.13 & 0.73 & 3.61 \\
  \hline
  \end{tabular}
}
\end{table}
\begin{table}
  \caption{Number and percentage of all changed lines}
  \label{tbl:all}
  \centering
  {\footnotesize\begin{tabular}{lrrrr}
  \hline
  & Lines & Mechanical & Manual & \textbf{Total} \\
  \hline
  lib/block\_diag.mli & 56 & 6 & 0 & 6 \\
  lib/block\_diag.ml & 58 & 9 & 0 & 9 \\
  lib/cov\_const.mli & 52 & 5 & 2 & 6 \\
  lib/cov\_const.ml & 141 & 14 & 10 & 16 \\
  lib/cov\_lin\_one.mli & 56 & 6 & 2 & 7 \\
  lib/cov\_lin\_one.ml & 149 & 17 & 13 & 26 \\
  lib/cov\_lin\_ard.mli & 56 & 6 & 2 & 7 \\
  lib/cov\_lin\_ard.ml & 188 & 32 & 9 & 39 \\
  lib/cov\_se\_iso.mli & 58 & 6 & 2 & 7 \\
  lib/cov\_se\_iso.ml & 343 & 71 & 9 & 78 \\
  lib/cov\_se\_fat.mli & 105 & 11 & 4 & 15 \\
  lib/cov\_se\_fat.ml & 680 & 174 & 30 & 199 \\
  lib/fitc\_gp.mli & 151 & 0 & 0 & 0 \\
  lib/fitc\_gp.ml & 2294 & 298 & 68 & 364 \\
  lib/interfaces.ml & 1008 & 197 & 26 & 215 \\
  lib/gpr\_utils.ml & 229 & 46 & 7 & 53 \\
  app/ocaml\_gpr.ml & 440 & 35 & 35 & 66 \\
  \hline
  \textbf{Total} & 6064 & 933 & 219 & 1113 \\
  \textbf{Percentage} & 100.00 & 15.39 & 3.61 & 18.35 \\
  \hline
  \end{tabular}
}
\end{table}

\section{Related work}
\label{sec:related-works}

Dependent ML (DML) \cite{DBLP:journals/jfp/Xi07} and sized
type \cite{DBLP:journals/lisp/ChinK01} can statically verify
consistency among the sizes of collections such as lists and arrays,
using dependent types on natural numbers.  In
ATS \cite{DBLP:conf/frocos/CuiDX05}, a successor of DML, BLAS and
LAPACK bindings are provided.  Advantages of dependent types over our
approach are: (1) they can represent more complex specifications
including inequalities such as array bounds; (2) they can verify the
consistency of sizes in the \emph{internal} implementations of vector
and matrix operations (though the BLAS and LAPACK bindings for ATS are
currently implemented as wrappers of C functions, so the internals are
not statically verified).  Conversely, our approach only requires
fairly standard ML types and module system, and application programs
can be ported almost mechanically, while dependent types generally
require non-trivial changes to the programming language and
application programs.

The dimensions of vectors and matrices can also be represented
\cite{hyone} using GADT, a lightweight form of dependent types.  
Existential types can be implemented not only using first-class modules
but also using GADT.

The idea of using phantom and generative types for static size checking is not novel.
Kiselyov and Shan \cite{DBLP:journals/entcs/KiselyovS07} implemented DML-like size checking (including inequalities,
e.g., array bound checking) by CPS encoding of existential types using
first-class polymorphism.
Eaton \cite{DBLP:conf/haskell/Eaton06} developed a linear algebra
library with static size checking for matrix operations as a
``strongly statically typed'' binding of GSLHaskell\footnote{
GSLHaskell is a binding of the GNU Scientific Library (GSL) \cite{GSL}, a library for
linear algebra and numerical computation on C and C++. GSL also provides interfaces for BLAS and LAPACK.}.
His basic idea is similar to ours, but he adopted Template
Haskell \cite{DBLP:journals/sigplan/SheardJ02} for the CPS encoding of generative types.
The approaches of \cite{DBLP:journals/entcs/KiselyovS07,DBLP:conf/haskell/Eaton06} need
CPS conversion when a generative type escapes its scope and thereby change the structures
of programs.
In contrast, we either implemented generative types with first-class modules
in OCaml instead of the CPS encoding, or else removed them in the first place,
like the conversions in EGPT.
Our contribution is the discovery that practical size checking for a linear algebra
library can be constructed on the simple idea of verifying mostly the equality of sizes
without significantly restructuring application programs.

Braibant and Pous \cite{DBLP:conf/itp/BraibantP10} implemented static
size checking of matrix operations using phantom types on Coq. It
requires more type annotations than our interface.

Eigen \cite{Eigen} is another practical linear algebra libraries with static size checking.
It does not statically check the consistency of dynamically determined sizes of matrices and vectors.

\section{Conclusions}
\label{sec:conclusions}

Our proposed linear algebra library interface SLAP uses generative phantom types
to statically ensure that most operations on matrices satisfy dimensional
constraints. It is based on a
simple idea---only the equality of sizes needs to be verified---and can be
realized by using a fairly standard type and module system of ML.  We
implemented this interface on top of LACAML and then ported
OCaml-GPR to it.  Most of the high-level matrix operations in the BLAS and LAPACK
linear algebra libraries were successfully typed,
and few non-trivial changes were required for the porting.

We did not find any bug in LACAML or OCaml-GPR, maybe because both
libraries have already been well tested and debugged or carefully
written in the first place.  However, in our experience of
implementing other (relatively small) programs\footnote{such as neural
  networks; see
  \url{https://github.com/akabe/slap/tree/master/examples}}, our
version of the libraries have been particularly useful when developing
a new library or application on top since they detect an error not
only \emph{earlier} (i.e., at compile time instead of runtime) but
also \emph{at higher level}: for instance, if the programmer misuses a
function of SGPR, an error is reported at the caller site rather than
somewhere deep inside the call stack from the function.

Interesting directions for future work include formalization of
the idea of generative phantom types, and extension of the static
types to enable verification of inequalities (in addition to
equalities), just to name a few.

\nocite{*}
\bibliographystyle{eptcs}
\bibliography{refs}

\begin{thebibliography}{10}
\providecommand{\bibitemdeclare}[2]{}
\providecommand{\surnamestart}{}
\providecommand{\surnameend}{}
\providecommand{\urlprefix}{Available at }
\providecommand{\url}[1]{\texttt{#1}}
\providecommand{\href}[2]{\texttt{#2}}
\providecommand{\urlalt}[2]{\href{#1}{#2}}
\providecommand{\doi}[1]{doi:\urlalt{http://dx.doi.org/#1}{#1}}
\providecommand{\bibinfo}[2]{#2}

\bibitemdeclare{article}{DBLP:journals/tcs/Blume01}
\bibitem{DBLP:journals/tcs/Blume01}
\bibinfo{author}{Matthias \surnamestart Blume\surnameend}
  (\bibinfo{year}{2001}): \emph{\bibinfo{title}{No-Longer-Foreign: Teaching an
  {ML} compiler to speak {C} "natively"}}.
\newblock {\sl \bibinfo{journal}{Electr. Notes Theor. Comput. Sci.}}
  \bibinfo{volume}{59}(\bibinfo{number}{1}), pp. \bibinfo{pages}{36--52}.
\newblock \urlprefix\url{http://dx.doi.org/10.1016/S1571-0661(05)80452-9}.

\bibitemdeclare{inproceedings}{DBLP:conf/itp/BraibantP10}
\bibitem{DBLP:conf/itp/BraibantP10}
\bibinfo{author}{Thomas \surnamestart Braibant\surnameend} \&
  \bibinfo{author}{Damien \surnamestart Pous\surnameend}
  (\bibinfo{year}{2010}): \emph{\bibinfo{title}{An Efficient {Coq} Tactic for
  Deciding Kleene Algebras}}.
\newblock In: {\sl \bibinfo{booktitle}{Interactive Theorem Proving, First
  International Conference, {ITP} 2010, Edinburgh, UK, July 11-14, 2010.
  Proceedings}}, {\sl \bibinfo{series}{Lecture Notes in Computer Science}}
  \bibinfo{volume}{6172}, \bibinfo{publisher}{Springer}, pp.
  \bibinfo{pages}{163--178}.
\newblock \urlprefix\url{http://dx.doi.org/10.1007/978-3-642-14052-5\_13}.

\bibitemdeclare{article}{DBLP:journals/lisp/ChinK01}
\bibitem{DBLP:journals/lisp/ChinK01}
\bibinfo{author}{Wei-Ngan \surnamestart Chin\surnameend} \&
  \bibinfo{author}{Siau-Cheng \surnamestart Khoo\surnameend}
  (\bibinfo{year}{2001}): \emph{\bibinfo{title}{Calculating Sized Types}}.
\newblock {\sl \bibinfo{journal}{Higher-Order and Symbolic Computation}}
  \bibinfo{volume}{14}(\bibinfo{number}{2-3}), pp. \bibinfo{pages}{261--300}.
\newblock \urlprefix\url{http://dx.doi.org/10.1023/A:1012996816178}.

\bibitemdeclare{inproceedings}{DBLP:conf/frocos/CuiDX05}
\bibitem{DBLP:conf/frocos/CuiDX05}
\bibinfo{author}{Sa~\surnamestart Cui\surnameend}, \bibinfo{author}{Kevin
  \surnamestart Donnelly\surnameend} \& \bibinfo{author}{Hongwei \surnamestart
  Xi\surnameend} (\bibinfo{year}{2005}): \emph{\bibinfo{title}{{ATS}: A
  Language That Combines Programming with Theorem Proving}}.
\newblock In: {\sl \bibinfo{booktitle}{FroCoS}}, {\sl \bibinfo{series}{Lecture
  Notes in Computer Science}} \bibinfo{volume}{3717},
  \bibinfo{publisher}{Springer}, pp. \bibinfo{pages}{310--320}.
\newblock \urlprefix\url{http://dx.doi.org/10.1007/11559306_19}.

\bibitemdeclare{article}{DBLP:journals/jfp/Danvy98}
\bibitem{DBLP:journals/jfp/Danvy98}
\bibinfo{author}{Olivier \surnamestart Danvy\surnameend}
  (\bibinfo{year}{1998}): \emph{\bibinfo{title}{Functional Unparsing}}.
\newblock {\sl \bibinfo{journal}{J. Funct. Program.}}
  \bibinfo{volume}{8}(\bibinfo{number}{6}), pp. \bibinfo{pages}{621--625}.
\newblock \urlprefix\url{http://dx.doi.org/10.1017/S0956796898003104}.

\bibitemdeclare{inproceedings}{DBLP:conf/haskell/Eaton06}
\bibitem{DBLP:conf/haskell/Eaton06}
\bibinfo{author}{Frederik \surnamestart Eaton\surnameend}
  (\bibinfo{year}{2006}): \emph{\bibinfo{title}{Statically typed linear algebra
  in Haskell}}.
\newblock In: {\sl \bibinfo{booktitle}{Proceedings of the {ACM} {SIGPLAN}
  Workshop on Haskell, Haskell 2006, Portland, Oregon, USA, September 17,
  2006}}, \bibinfo{publisher}{{ACM}}, pp. \bibinfo{pages}{120--121}.
\newblock \urlprefix\url{http://dx.doi.org/10.1145/1159842.1159859}.

\bibitemdeclare{misc}{Eigen}
\bibitem{Eigen}
\emph{\bibinfo{title}{{E}igen}}.
\newblock \bibinfo{howpublished}{\url{http://eigen.tuxfamily.org/}}.

\bibitemdeclare{article}{DBLP:journals/jfp/FluetP06}
\bibitem{DBLP:journals/jfp/FluetP06}
\bibinfo{author}{Matthew \surnamestart Fluet\surnameend} \&
  \bibinfo{author}{Riccardo \surnamestart Pucella\surnameend}
  (\bibinfo{year}{2006}): \emph{\bibinfo{title}{Phantom types and subtyping}}.
\newblock {\sl \bibinfo{journal}{J. Funct. Program.}}
  \bibinfo{volume}{16}(\bibinfo{number}{6}), pp. \bibinfo{pages}{751--791}.
\newblock \urlprefix\url{http://dx.doi.org/10.1017/S0956796806006046}.

\bibitemdeclare{misc}{GSL}
\bibitem{GSL}
\bibinfo{author}{Mark \surnamestart Galassi\surnameend} et~al.:
  \emph{\bibinfo{title}{the {GNU} Scientific Library ({GSL})}}.
\newblock \bibinfo{howpublished}{\url{http://www.gnu.org/software/gsl/}}.

\bibitemdeclare{incollection}{Hin03Fun}
\bibitem{Hin03Fun}
\bibinfo{author}{Ralf \surnamestart Hinze\surnameend} (\bibinfo{year}{2003}):
  \emph{\bibinfo{title}{Fun with phantom types}}.
\newblock In \bibinfo{editor}{Jeremy \surnamestart Gibbons\surnameend} \&
  \bibinfo{editor}{Oege \surnamestart {de Moor}\surnameend}, editors: {\sl
  \bibinfo{booktitle}{The Fun of Programming}}, \bibinfo{series}{Cornerstones
  of Computing}, \bibinfo{publisher}{Palgrave Macmillan}, pp.
  \bibinfo{pages}{245--262}.

\bibitemdeclare{misc}{hyone}
\bibitem{hyone}
\bibinfo{author}{\surnamestart hyone\surnameend}: \emph{\bibinfo{title}{Length
  Indexed Matrix and Indexed Functor}}.
\newblock \bibinfo{howpublished}{\url{https://gist.github.com/hyone/3990929}}.

\bibitemdeclare{article}{DBLP:journals/entcs/KiselyovS07}
\bibitem{DBLP:journals/entcs/KiselyovS07}
\bibinfo{author}{Oleg \surnamestart Kiselyov\surnameend} \&
  \bibinfo{author}{Chung \surnamestart chieh Shan\surnameend}
  (\bibinfo{year}{2007}): \emph{\bibinfo{title}{Lightweight Static
  Capabilities}}.
\newblock {\sl \bibinfo{journal}{Electr. Notes Theor. Comput. Sci.}}
  \bibinfo{volume}{174}(\bibinfo{number}{7}), pp. \bibinfo{pages}{79--104}.
\newblock \urlprefix\url{http://dx.doi.org/10.1016/j.entcs.2006.10.039}.

\bibitemdeclare{inproceedings}{DBLP:conf/dsl/LeijenM99}
\bibitem{DBLP:conf/dsl/LeijenM99}
\bibinfo{author}{Daan \surnamestart Leijen\surnameend} \& \bibinfo{author}{Erik
  \surnamestart Meijer\surnameend} (\bibinfo{year}{1999}):
  \emph{\bibinfo{title}{Domain specific embedded compilers}}.
\newblock In: {\sl \bibinfo{booktitle}{Proceedings of the Second Conference on
  Domain-Specific Languages {(DSL} '99), Austin, Texas, USA, October 3-5,
  1999}}, \bibinfo{publisher}{{ACM}}, pp. \bibinfo{pages}{109--122}.
\newblock \urlprefix\url{http://dx.doi.org/10.1145/331960.331977}.

\bibitemdeclare{misc}{OCaml-GPR}
\bibitem{OCaml-GPR}
\bibinfo{author}{Markus \surnamestart Mottl\surnameend}:
  \emph{\bibinfo{title}{{OCaml-GPR} -- Efficient {G}aussian Process Regression
  in {OCaml}}}.
\newblock \bibinfo{howpublished}{\url{https://github.com/mmottl/gpr}}.

\bibitemdeclare{misc}{Lacaml}
\bibitem{Lacaml}
\bibinfo{author}{Markus \surnamestart Mottl\surnameend} \&
  \bibinfo{author}{Christophe \surnamestart Troestler\surnameend}:
  \emph{\bibinfo{title}{{LACAML} -- Linear Algebra for {OCaml}}}.
\newblock \bibinfo{howpublished}{\url{https://github.com/mmottl/lacaml}}.

\bibitemdeclare{misc}{BLAS}
\bibitem{BLAS}
\bibinfo{author}{\surnamestart {NetLib}\surnameend}:
  \emph{\bibinfo{title}{{BLAS} ({B}asic {L}inear {A}lgebra {S}ubprograms)}}.
\newblock \bibinfo{howpublished}{\url{http://www.netlib.org/blas/}}.

\bibitemdeclare{misc}{LAPACK}
\bibitem{LAPACK}
\bibinfo{author}{\surnamestart {NetLib}\surnameend}:
  \emph{\bibinfo{title}{{LAPACK} -- {L}inear {A}lgebra {PACK}age}}.
\newblock \bibinfo{howpublished}{\url{http://www.netlib.org/lapack/}}.

\bibitemdeclare{misc}{OptiML}
\bibitem{OptiML}
\bibinfo{author}{\surnamestart {S}tanford {U}niversity's {P}ervasive
  {P}arallelism~{L}aboratory ({PPL})\surnameend}:
  \emph{\bibinfo{title}{{O}pti{ML}}}.
\newblock
  \bibinfo{howpublished}{\url{http://stanford-ppl.github.io/Delite/optiml/}}.

\bibitemdeclare{article}{DBLP:journals/sigplan/SheardJ02}
\bibitem{DBLP:journals/sigplan/SheardJ02}
\bibinfo{author}{Tim \surnamestart Sheard\surnameend} \& \bibinfo{author}{Simon
  L.~Peyton \surnamestart Jones\surnameend} (\bibinfo{year}{2002}):
  \emph{\bibinfo{title}{Template meta-programming for Haskell}}.
\newblock {\sl \bibinfo{journal}{{SIGPLAN} Notices}}
  \bibinfo{volume}{37}(\bibinfo{number}{12}), pp. \bibinfo{pages}{60--75}.
\newblock \urlprefix\url{http://dx.doi.org/10.1145/636517.636528}.

\bibitemdeclare{inproceedings}{DBLP:conf/icml/SujeethLBRCWAOO11}
\bibitem{DBLP:conf/icml/SujeethLBRCWAOO11}
\bibinfo{author}{Arvind~K. \surnamestart Sujeeth\surnameend},
  \bibinfo{author}{HyoukJoong \surnamestart Lee\surnameend},
  \bibinfo{author}{Kevin~J. \surnamestart Brown\surnameend},
  \bibinfo{author}{Tiark \surnamestart Rompf\surnameend},
  \bibinfo{author}{Hassan \surnamestart Chafi\surnameend},
  \bibinfo{author}{Michael \surnamestart Wu\surnameend},
  \bibinfo{author}{Anand~R. \surnamestart Atreya\surnameend},
  \bibinfo{author}{Martin \surnamestart Odersky\surnameend} \&
  \bibinfo{author}{Kunle \surnamestart Olukotun\surnameend}
  (\bibinfo{year}{2011}): \emph{\bibinfo{title}{{O}pti{ML}: An Implicitly
  Parallel Domain-Specific Language for Machine Learning}}.
\newblock In: {\sl \bibinfo{booktitle}{Proceedings of the 28th International
  Conference on Machine Learning, {ICML} 2011, Bellevue, Washington, USA, June
  28 - July 2, 2011}}, \bibinfo{publisher}{Omnipress}, pp.
  \bibinfo{pages}{609--616}.

\bibitemdeclare{misc}{uBlas}
\bibitem{uBlas}
\emph{\bibinfo{title}{u{B}las}}.
\newblock
  \bibinfo{howpublished}{\url{http://www.boost.org/doc/libs/1_55_0/libs/numeric/ublas/doc/}}.

\bibitemdeclare{article}{DBLP:journals/jfp/Xi07}
\bibitem{DBLP:journals/jfp/Xi07}
\bibinfo{author}{Hongwei \surnamestart Xi\surnameend} (\bibinfo{year}{2007}):
  \emph{\bibinfo{title}{Dependent {ML} -- An approach to practical programming
  with dependent types}}.
\newblock {\sl \bibinfo{journal}{J. Funct. Program.}}
  \bibinfo{volume}{17}(\bibinfo{number}{2}), pp. \bibinfo{pages}{215--286}.
\newblock \urlprefix\url{http://dx.doi.org/10.1017/S0956796806006216}.

\end{thebibliography}

\appendix

\section{Encoding of subtyping}
\label{subtyping}

We explain generalization of the subtyping encoding used in Section \ref{discrete}.
To start with, consider a subtyping with only one base case \lstinline|T :> U|.
Table \ref{tab:subtyping1} shows encoding of the types \lstinline|T| and \lstinline|U|
(which depends on their positions of appearance).
\begin{table}
  \centering
  \caption{Encoding of supertype \lstinline|T| and subtype \lstinline|U|}
  \begin{tabular}{ccc}
    \hline
    & Positive (covariant) position & Negative (contravariant) position \\
    \hline
    Supertype \lstinline|T| &
    \lstinline| t      $\tau$| &
    \lstinline|'t_or_u $\tau$| \\
    Subtype \lstinline|U| &
    \lstinline|'t_or_u $\tau$| &
    \lstinline|      u $\tau$| \\
    \hline
  \end{tabular}
  \label{tab:subtyping1}
\end{table}
It is similar to the subtyping encoding for the types of contiguous and discrete
matrices (without type parameters \lstinline|'m| and \lstinline|'n| for dimensions)
in Section \ref{discrete}.
\lstinline|t| and \lstinline|u| are phantom types, and
\lstinline|'t_or_u $\tau$| is the sum type of \lstinline|T| and \lstinline|U| where
\lstinline|'t_or_u| is a phantom type parameter.

\begin{figure}
  \centering
  \includegraphics{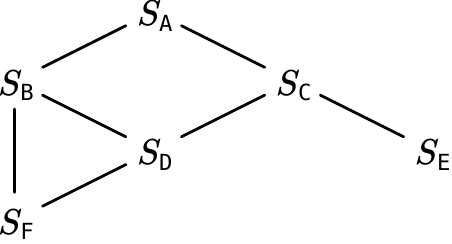}
  \caption{Example of powerset lattice}
  \label{fig:powerset-lattice}
\end{figure}

More generally, to encode an arbitrary subtyping hierarchy with a
finite number of base cases, we give an encoding of \emph{powerset
  lattices}.  The powerset lattice of a finite set $S$ is the lattice
of all subsets of $S$, ordered by inclusion. As in
\cite{DBLP:journals/jfp/FluetP06}, we represent our subtyping relation
as the inclusion relation between subsets of some $S$ so that
\lstinline|T :> U| iff $S_{\Type{T}} \supseteq S_{\Type{U}}$ for any
types \lstinline|T| and \lstinline|U|, where $S_{\Type{T}}$ and
$S_{\Type{U}}$ are some appropriate subsets of $S$ corresponding to
\lstinline|T| and \lstinline|U|, respectively.  For example, consider
the subtyping relation illustrated in Figure \ref{fig:powerset-lattice}, where
\lstinline|A| is the largest type, and \lstinline|E| and \lstinline|F|
are the smallest.  Let $S$ be $\{1,2,3,4\}$.  Then we can take, e.g.,
$S_{\Type{A}} = \{1,2,3,4\}$, $S_{\Type{B}} = \{1,2\}$, $S_{\Type{C}}
= \{1,3,4\}$, $S_{\Type{D}} = \{2,3\}$, $S_{\Type{E}} = \{4\}$, and
$S_{\Type{F}} = \{1\}$.

We now assume a total ordering $s_1, s_2, \dots, s_n$ among the elements
of $S$, where $n$ is the cardinality of $S$. A base type \lstinline|T| is
encoded as $\langle \Type{T} \rangle_+\ \tau$ at covariant positions,
and as $\langle \Type{T} \rangle_-\ \tau$ at contravariant positions,
where $\langle \Type{T} \rangle_+$ and $\langle \Type{T} \rangle_-$ are
$n$-tuple types defined as
\begin{align*}
  \langle \Type{T} \rangle_+ \quad
  & \defeq \quad
  t_1  ~\Type{*}~ \cdots ~\Type{*}~ t_n \quad
  \text{where}~~
  t_i =
  \begin{cases}
    \Type{w} & \text{if $s_i \in S_{\Type{T}}$} \\
    \Type{'a}_i & \text{otherwise},
  \end{cases} \\
  \langle \Type{T} \rangle_- \quad
  & \defeq \quad
  t_1  ~\Type{*}~ \cdots ~\Type{*}~ t_n \quad
  \text{where}~~
  t_i =
  \begin{cases}
    \Type{'a}_i & \text{if $s_i \in S_{\Type{T}}$} \\
    \Type{z} & \text{otherwise},
  \end{cases}
\end{align*}
for phantom types \lstinline|w| and \lstinline|z|.
We require that every type parameter \lstinline|'a$_i$| is fresh
in \emph{each} $\langle \cdot \rangle_+$ and $\langle \cdot \rangle_-$.

\begin{table}
  \centering
  \caption{Encoding of the subtyping hierarchy of Figure \ref{fig:powerset-lattice}}
  \begin{tabular}{c@{\hskip2\bigskipamount}c@{\hskip2\bigskipamount}c}
    \hline
    & $\langle \cdot \rangle_+$ & $\langle \cdot \rangle_-$ \\
    \hline
    \lstinline|A| &
    \lstinline| w$_{\phantom{1}}$ *  w$_{\phantom{2}}$ *  w$_{\phantom{3}}$ *  w$_{\phantom{4}}$| &
    \lstinline|'a$_1$ * 'a$_2$ * 'a$_3$ * 'a$_4$| \\
    \lstinline|B| &
    \lstinline| w$_{\phantom{1}}$ *  w$_{\phantom{2}}$ * 'a$_3$ * 'a$_4$| &
    \lstinline|'a$_1$ * 'a$_2$ *  z$_{\phantom{3}}$ *  z$_{\phantom{4}}$| \\
    \lstinline|C| &
    \lstinline| w$_{\phantom{1}}$ * 'a$_2$ *  w$_{\phantom{3}}$ *  w$_{\phantom{4}}$| &
    \lstinline|'a$_1$ *  z$_{\phantom{2}}$ * 'a$_3$ * 'a$_4$| \\
    \lstinline|D| &
    \lstinline| w$_{\phantom{1}}$ * 'a$_2$ *  w$_{\phantom{3}}$ * 'a$_4$| &
    \lstinline|'a$_1$ *  z$_{\phantom{2}}$ * 'a$_3$ *  z$_{\phantom{4}}$| \\
    \lstinline|E| &
    \lstinline|'a$_1$ * 'a$_2$ * 'a$_3$ *  w$_{\phantom{4}}$| &
    \lstinline| z$_{\phantom{1}}$ *  z$_{\phantom{2}}$ *  z$_{\phantom{3}}$ * 'a$_4$| \\
    \lstinline|F| &
    \lstinline| w$_{\phantom{1}}$ * 'a$_2$ * 'a$_3$ * 'a$_4$| &
    \lstinline|'a$_1$ *  z$_{\phantom{3}}$ *  z$_{\phantom{4}}$ *  z$_{\phantom{5}}$| \\
    \hline
  \end{tabular}
  \label{tab:subtyping2}
\end{table}
Table \ref{tab:subtyping2} shows the encoding of the subtyping relation in
Figure \ref{fig:powerset-lattice}. We can verify, e.g., that
$\langle \Type{E} \rangle_+$ can be unified with $\langle \Type{A} \rangle_-$,
$\langle \Type{C} \rangle_-$, or $\langle \Type{E} \rangle_-$,
but not with $\langle \Type{B} \rangle_-$, $\langle \Type{D} \rangle_-$, or
$\langle \Type{F} \rangle_-$.  That is, a value of
type \lstinline|E| can be passed to a function as an argument of type
\lstinline|A|, \lstinline|C|, or \lstinline|E|, but not as \lstinline|B|,
\lstinline|D|, or \lstinline|F|.
In addition, if we put values of type \lstinline|E| and \lstinline|F| in the same list,
it can be passed as an argument of type \lstinline|A list| or \lstinline|C list|
(i.e., a list of elements with a common supertype of \lstinline|E| and \lstinline|F|).

We implement \lstinline|Bot|, a subtype of any type,
as a single type parameter (i.e., $\forall \alpha.~\alpha$), which can be instantiated
to any type. By replacing \lstinline|'a$_i$| with \lstinline|Bot|
in the definitions of $\langle \cdot \rangle_+$ and $\langle \cdot \rangle_-$,
we obtain $\mbox{\lstinline|U <: T|} \iff
\langle \Type{U} \rangle_+~\Type{<:}~\langle \Type{T} \rangle_+ \land
\langle \Type{U} \rangle_-~\Type{:>}~\langle \Type{T} \rangle_-$.

\paragraph{Related work}
Fluet and Pucella \cite{DBLP:journals/jfp/FluetP06} also proposed a subtyping scheme using phantom types
on the ML type system.
They focused on an encoding of Hindley-Milner polymorphism extended with subtyping.
Their approach can encode a type like $\forall \alpha\Type{\,<:\,T}.~\alpha \to \alpha$,
but does not achieve the contravariance of argument types.
In contrast, our approach accomplishes both covariance and contravariance
while it does not support universal types.

\section{Generalization of flag-dependent function types}

We generalize the phantom type trick in Section \ref{sec:funtype} for function types that
depend on flags (cf. Danvy's typing of \lstinline|printf| \cite{DBLP:journals/jfp/Danvy98}).
The type of the following function depends on values of
$\texttt{x}_1, \dots, \texttt{x}_n$ ($n = 1$ in many functions of BLAS and LAPACK,
but several functions such as \lstinline|gemm| and \lstinline|gesvd| take two
or more flags).
\begin{lstlisting}
val f : $\Pi$x$_1$:t$_1$. $\Pi$x$_2$:t$_2$. $\dots$ $\Pi$x$_n$:t$_n$. ($\mathcal{T}_1($x$_1)$, $\mathcal{T}_2($x$_2)$, $\dots$, $\mathcal{T}_n($x$_n)$) u
\end{lstlisting}
$\mathcal{T}_i$ is a function that maps a flag value to an ML type,
\lstinline|($\alpha_1$, $\dots$, $\alpha_n$) u| is an ML type,
and \lstinline|t$_i$| is the type of the $i$th flags, e.g., \lstinline/[`N | `T | `C]/ (for
transpose flags), \lstinline/[`L | `R]/ (for side flags) or
\lstinline/[`A | `S | `O | `N]/ (for SVD job flags).
We assume that $\mathcal{T}_i$ does not contain dependent types.

We consider how to type \lstinline|f| without dependent types.
First, we represent each type \lstinline|t$_i$| and
each flag \lstinline|v$_{ij}$| of type \lstinline|t$_i$| as follows:
\begin{lstlisting}
type $\alpha$ tt$_i$ (* = t$_i$ *)
val tt$_i$_v$_{ij}$ : $\mathcal{T}_i($v$_{ij})$ tt$_i$ (* = v$_{ij}$ *)
\end{lstlisting}
Then \lstinline|f| takes the flag representations \lstinline|tt$_i$_v$_{ij}$| as arguments of types \lstinline|$\alpha_i$ tt$_i$|, thereby receiving
the types \lstinline|$\mathcal{T}_i($v$_{ij})$ tt$_i$| as $\alpha_i$:
\begin{lstlisting}
val f : $\forall \alpha_1, \alpha_2, \dots, \alpha_n$. $\alpha_1$ tt$_1$ $\to$ $\alpha_2$ tt$_2$ $\to \dots \to$ $\alpha_n$ tt$_n$ $\to$ ($\alpha_1$, $\alpha_2$, $\dots$, $\alpha_n$) u
\end{lstlisting}

We show the encoding of the type of \lstinline|gemm| as an example.
The original type of it is
\begin{lstlisting}
val gemm : $\Pi$transa:[`N | `T | `C] $\to$ $\Pi$transb:[`N | `T | `C] $\to$
           ($\mathcal{T}($transa$)$, $\mathcal{T}($transb$)$) u
\end{lstlisting}
where
\begin{lstlisting}
(('am, 'ak) mat $\to$ ('m, 'k) mat, ('bk, 'bn) mat $\to$ ('k, 'n) mat) u
  $=$ ?beta:num_type $\to$ ?c:('m, 'n) mat (* $\bm{C}$ *) $\to$
      ?alpha:num_type $\to$ ('am, 'ak) mat (* $\bm{A}$ *) $\to$
      ('bk, 'bn) mat (* $\bm{B}$ *) $\to$ ('m, 'n) mat (* $\bm{C}$ *)
\end{lstlisting}
and
\[
  \mathcal{T}(\verb|trans|) =
  \begin{cases}
    \verb|('m, 'n) mat| \to \verb|('m, 'n) mat| & (\verb|trans| = \verb|`N|) \\
    \verb|('m, 'n) mat| \to \verb|('n, 'm) mat| & (\verb|trans| = \verb|`T|, \verb|`C|).
  \end{cases}
\]
Our representations of the flags are
\begin{lstlisting}
type $\alpha$ trans (* = [`N | `T | `C] *)
val normal : $\mathcal{T}($`N$)$ trans (* = `N *)
val trans  : $\mathcal{T}($`T$)$ trans (* = `T *)
val conjtr : $\mathcal{T}($`C$)$ trans (* = `C *)
\end{lstlisting}
and our type of \lstinline|gemm| is:
\begin{lstlisting}
val gemm : $\forall \alpha_1, \alpha_2$. $\alpha_1$ trans $\to$ $\alpha_2$ trans $\to$ ($\alpha_1$, $\alpha_2$) u
\end{lstlisting}
Side flags and SVD job flags can be represented similarly.

\end{document}